\documentclass[aps,twocolumn,superscriptaddress,showpacs,preprintnumbers,nofootinbib,amsmath,amssymb,]{revtex4}
\usepackage{braket}
\usepackage{bm}
\usepackage{dcolumn}
\usepackage{hyperref}
\usepackage[dvipdfmx,final]{graphicx}
\usepackage{ulem}
\input{colordvi.tex}
\usepackage{color}
\newcommand{\Mpc}{\,{\rm Mpc}}
\newcommand{\be}{\begin{eqnarray}}
\newcommand{\ee}{\end{eqnarray}}

\begin{document}
\title{Early recombination as a solution to the $H_0$ tension}

\author{Toyokazu Sekiguchi}
\email{tsekiguc@post.kek.jp}
\affiliation{Theory Center, IPNS, KEK, Tsukuba 305-0801, Japan}

\author{Tomo Takahashi}
\email{tomot@cc.saga-u.ac.jp}
\affiliation{Department of Physics, Saga University, Saga 840-8502, Japan}
\date{\today}

\begin{abstract}
We show that the $H_0$ tension can be resolved by making recombination earlier, keeping the fit to cosmic microwave background (CMB) data almost intact.
We provide a suite of general necessary conditions to give a good fit to CMB data while realizing a high value of $H_0$ suggested by local measurements.
As a concrete example for a successful scenario with early recombination, we demonstrate that a model with time-varying $m_e$ can indeed satisfy all the conditions. 
We further show that such a model can also be well fitted to low-$z$ distance measurements of baryon acoustic oscillation (BAO) and type-Ia supernovae (SNeIa) 
with a simple extension of the model. Time-varying $m_e$ in the framework of $\Omega_k\Lambda$CDM is found to be a sufficient and excellent example as 
a solution to the $H_0$ tension, yielding $H_0=72.3_{-2.8} ^{+2.7}$\,km/sec/Mpc from the combination of CMB, BAO and SNeIa data {\it even  without} 
incorporating any direct local $H_0$ measurements. 
Apart from the $H_0$ tension, this model is also favored from the viewpoint of the CMB lensing anomaly.

\end{abstract}

\preprint{
KEK-TH-2238\quad
KEK-Cosmo-???
}


\maketitle

{\it Introduction.}---
The Hubble-Lema\^itre constant $H_0$ is one of the most relevant cosmological parameters characterizing the 
Universe. It has long been studied by the distance ladder, 
which now utilizes Cepheids and type Ia supernovae (SNeIa) as standard candles \cite{Riess:2019cxk}.
Meanwhile, many other means have been devised.
For instance, gravitational lens time-delay now rivals the distance ladder in local (almost direct) measurements of $H_0$ \cite{Wong:2019kwg}.
Moreover, the cosmic microwave background (CMB) and the baryon acoustic oscillations (BAO), allow us to measure 
cosmic distances to very different redshifts ($z\simeq 10^3$ and $z\lesssim2$) based on the scales of the sound horizon of photon-baryon fluid, $r_s$, at recombination $z=z_*$ and the drag epoch $z=z_{\rm drag}$, respectively. 
Consistency in the cosmic expansion history over such a huge range of redshifts enables us to infer $H_0$. 

However, as measurements of $H_0$ become more precise, 
disagreements grow apparent between local direct measurements and 
other indirect ones such as CMB. The value of $H_0$ from local measurements, 
$H_0=(73.8\pm1.0)\,{\rm km/sec/Mpc}$ \cite{Riess:2020sih}, is about $10\%$ larger than that from CMB, 
$H_0 = (67.36\pm0.54)\,{\rm km/sec/Mpc}$ \cite{Aghanim:2018eyx}, 
assuming the canonical flat $\Lambda$CDM ($\Lambda$CDM hereafter) model. 
Significance of the $H_0$ tension is now more than $5\sigma$.
Interestingly, different and independent measurements appear consistent within either local or indirect measurements
(For recent review, see \cite{Riess:2020sih}). 
This indicates that single systematic error alone cannot remove the tension.

A number of cosmological solutions have been proposed already.
However, it seems extremely difficult to solve the tension when one combines
various observations such as CMB, BAO and SNeIa.
The reason for the difficulty has been clarified in 
\cite{Aylor:2018drw,Knox:2019rjx}.
SNeIa and distance ladder jointly measure luminosity distance seamlessly at $z\lesssim 2$.
This gives the transverse distance $D_M(z)$ at the redshifts of BAO measurements, where $D_M(z)$ is given by
\be
D_M(z) = \begin{cases}
\displaystyle\frac {\sin\left[\sqrt{-\Omega_k}H_0\chi(z)\right]}
{\sqrt{-\Omega_k}H_0} & \mbox{for~}\Omega_k<0\mbox{~(closed)} \\ \\
\chi(z) & \mbox{for~}\Omega_k=0\mbox{~(flat)}\\  \\
\displaystyle\frac 
{\sinh\left[\sqrt{\Omega_k}H_0\chi(z)\right]}
{\sqrt{\Omega_k}H_0} & 
\mbox{for~}\Omega_k>0\mbox{~(open)}
\end{cases},\label{eq:DM}
\ee
with $ \chi(z)=\int^z_0\frac{dz}{H(z)}$ being the comoving distance to $z$. 
This enables model-independent estimation of  $r_s(z_{\rm drag})$. 
Enhancing $H_0$ by 10\% requires decreasing $r_s(z_*)\propto r_s(z_{\rm drag})$ by the same rate,\footnote{
Given the baryon drag at $z_*$, $R(z_*)=3\rho_b(z_*)/4\rho_\gamma(z_*)$, 
which is very precisely determined by CMB power spectra, 
specifying either of $r_s(z_*)$ or $r_s(z_{\rm drag})$  virtually 
determines the other.}
which is very difficult keeping a reasonable fit to CMB. This also explains why models modifying 
only late-time expansion can increase $H_0$ only marginally.

Considerations above lead to following four necessary conditions
which successful cosmological solutions to the $H_0$ tension should satisfy: 
\begin{enumerate}
\item 
In order not to spoil the successful fit achieved by $\Lambda$CDM,
CMB power spectra should be left almost intact except at low-$\ell$, where cosmic variance is large.
\item 
$r_s(z_*) \propto r_s(z_{\rm drag})$ is reduced by $\simeq 10$\%.
\item 
$D_M(z_*)$ is reduced, so that $\theta_s (z_\ast)= r_s(z_*)/D_M(z_*)$ is kept constant (this is somewhat redundant with the condition 1).
\item BAO, SNeIa, and other low-$z$ distance measurements should be satisfied.
\end{enumerate}

With the first condition being met, the second condition is quite
difficult to be satisfied. Many attempts have tried to increase the expansion rate by e.g. 
adding extra energy components (See \cite{Knox:2019rjx} for review). 
However, these modifications have some limitations since the relative scale between the 
sound horizon and the Silk scale, or the photon diffusion length, also varies,
which inevitably violates the first condition \cite{Knox:2019rjx}.
This is the reason why those attempts
can mitigate the $H_0$ tension only somewhat partially.

We in this {\it Letter} pursue a cosmological solution to the $H_0$ tension, 
in particular focusing on modified recombination (See earlier studies \cite{Chiang:2018xpn,Liu:2019awo} but without concrete models).
We first argue how one can modify recombination epoch keeping CMB power spectra almost unchanged.
Then as a working example, we discuss a model with time-varying electron mass $m_e$
(for possible models of time-varying $m_e$, see, e.g., a recent review \cite{Martins:2017yxk})\footnote{
In the following discussion, we assume a different value of $m_e$ for the present time and the recombination epoch.
In this sense, $m_e$ is time-varying, however, we make a simplified assumption where $m_e$ is constant until some time after 
recombination, then at some epoch, $m_e$ becomes the present value.
}, which can shift $z_*$ and $z_{\rm drag}$ from the  baseline model sizeably without affecting CMB power spectra much. 

In the following, we often refer to the Planck 2018 best-fit 
$\Lambda$CDM model \cite{Aghanim:2018eyx} as the baseline.
The reduced Hubble constant and density parameters are given by
$h=H_0\,{\rm [100\,km/sec/Mpc]}$ and e.g. $\omega_i = \Omega_i h^2$ for component $i$, respectively.
Let $\Delta_x$ denote the fractional variation in a quantity $x$ from the baseline value 
[e.g. $\Delta_{m_e}=\log(m_e/m_{e,{\rm baseline}})$]. 

{\it Effects of early recombination on CMB.---}
Let us discuss effects of early recombination on CMB and how to cancel those effects 
by varying cosmological parameters.
In the analytical argument below, we utilize the scale factor at recombination $a=a_\ast$, which is 
useful since it can well capture effects of modified recombination on CMB.

CMB observations tightly constrain the following two quantities at the recombination $a=a_*$:
\be
R(x)&=&\frac{3\omega_b a_*}{4\omega_\gamma}x, \label{eq:R}\\
\left[a^2H\right](x)&=&\frac1{L}\sqrt{\omega_ma_*x+\omega_r}, \label{eq:dtauda}
\ee
where $x\equiv a/a_*$ is the scale factor normalized to unity at recombination
and $L=(H_0/h)^{-1} \simeq 2998\Mpc$ is a constant length.
The former gives the baryon drag, which is measured by the relative heights of even and odd acoustic peaks.
The latter determines the early integrated Sachs-Wolfe (ISW) effect, which is measured by the heights of acoustic peaks relative to the SW plateau. 
From Eqs.~\eqref{eq:R} and \eqref{eq:dtauda}, we can leave both $R$ and $a^2H$ unaffected as functions of $x$
by varying $\omega_b$ and $\omega_m$ inversely proportionally to $a_*$:
\be
\Delta_{\omega_b}=\Delta_{\omega_m}=-\Delta_{a_*}.\label{eq:sim1}
\ee

Now we consider the sound horizon at the recombination epoch:
\be
\label{eq:r_s}
r_s(z_*)&=&\frac{a_*}{\sqrt3}\int^1_0 \frac1{\sqrt{1+R(x)}} \frac{dx}{[a^2H](x)},
\ee
from which 
we can immediately see $r_s(z_*)\propto a_*$ when
we vary $\omega_b$ and $\omega_m$ in accord with Eq.~\eqref{eq:sim1} (i.e., $R$ and $a^2H$ remain unchanged as functions of $x$).
Not to change CMB power spectra, the relative scale of the Silk scale, $1/d_{D*}$, to $r_{s*}$ should be kept unchanged, where
\be
\frac1{k_D(z_*)^2}&=&\frac{a_*^2}6\int^1_0 \frac{R^2+\frac{16}{15}(1+R)}{(1+R)^2} 
\frac{1}{a_*^2 n_e \sigma_T}
\frac{dx/x}{[a^2H]}. \label{eq:defkD}
\ee
This requires
\be
1/k_D(z_*)\propto a_*. \label{eq:kD}
\ee

Finally, the viewing angle of the sound horizon, $r_s(z_\ast)/D_M (z_\ast)$, should be kept constant, which
means $D_M(z_*)$ should vary proportionally to $a_*$.
Within $\Lambda$CDM background,  
we find that
\be
\Delta_h \approx - 3.23 {\Delta_{a_\ast}} 
\label{eq:sim2}
\ee
approximately realizes $D_M(z_*)\propto a_*$, where the numerical coefficient 
is evaluated around the baseline. 

Conditions \eqref{eq:sim1} and \eqref{eq:sim2} can be easily satisfied by varying standard cosmological parameters.
Contrastively, Eq.~\eqref{eq:kD} is non-trivial. As we will show below, 
varying $m_e$ models can satisfy this non-trivial condition while the other ones are also held.

{\it Varying $m_e$ and CMB power spectra.---}
As a working example, here we consider a model with time-varying $m_e$. 
The electron mass $m_e$ affects physics of CMB at recombination through the following aspects:
\begin{itemize}
\item Energy levels of hydrogen: $E\propto m_e$
\item Thomson scattering cross section: $\sigma_{\rm T}\propto m_e^{-2}$
\item Others (two-photon decay rate, photo-ionization cross section, recombination coefficients etc.)
\end{itemize}
If recombination proceeds in thermal equilibrium, the third effects can be omitted. 
Although non-equilibrium processes are evident in observed CMB power spectra, 
their impact is indeed relatively minor as long as $m_e$ alone is varied \cite{Ade:2014zfo}. 
Neglecting the third effects to simplify discussion, 
$a_*$ is inversely proportional to $m_e$ 
through the first effect:
\be
\Delta_{m_e}=\Delta_{T_\gamma(z_*)} =-\Delta_{a_*}. \label{eq:me}
\ee

\begin{figure*}[!t]
  \makebox[\textwidth][c]{\includegraphics[width=1.15\textwidth]{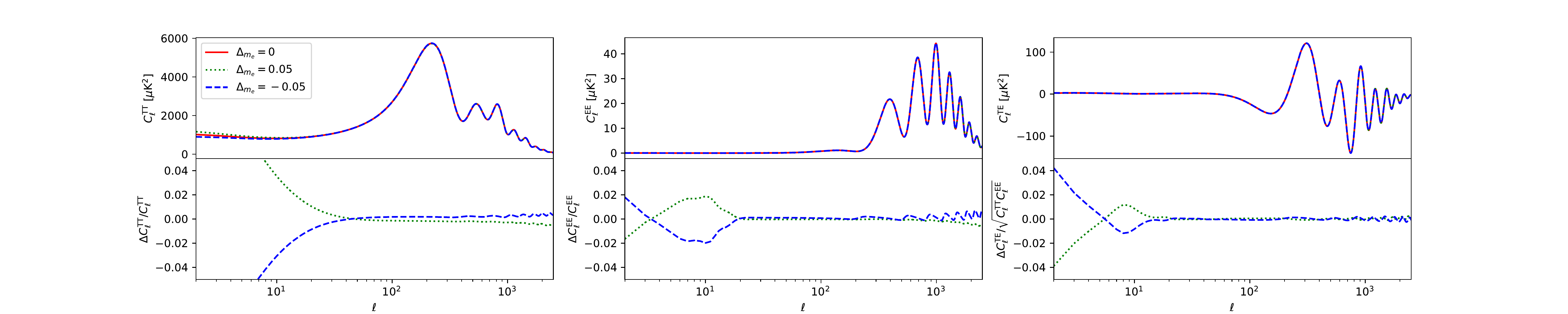}}
\caption{
\label{fig:cell}
CMB power spectra with $\Delta_{m_e}=0,\,\pm 0.05$ in $\Lambda$CDM background along the parameter direction
Eqs. \eqref{eq:sim1}, \eqref{eq:sim2} and  \eqref{eq:me}.
}
\end{figure*}

To see how $1/k_D$ in Eq. \eqref{eq:defkD} is modified in response to $m_e$,
let us consider the following factor:
\be
a_*^2n_e\sigma_T&=&x_e \frac{1-Y_p}{m_H} \frac{\rho_{\rm crit}}{h^2}(\omega_ba_*)\left(\frac{\sigma_T}{a_*^2}\right)\frac1{x^3},
\label{eq:f}
\ee
where $x_e$, $Y_p$, $m_H$ and $\rho_{\rm crit}=\frac{3h^2}{8\pi L^2G}$
are the ionization fraction, the mass fraction of $^4$He, the hydrogen mass, 
and the critical density, respectively.
When we vary $\omega_b$ according to Eq.~\eqref{eq:sim1},
Eq.~\eqref{eq:f} does not change as a function of $x$.\footnote{
If we neglect the non-equilibrium nature of recombination, $x_e$ does not change  as function of $x$. 
We also have omitted the marginal dependence of $Y_p$ on $\omega_b$ in the Big Bang nucleosynthesis prediction.
}
Thus, the integral in Eq. \eqref{eq:defkD} is kept constant, which means 
$1/k_D(a_\ast) \propto a_\ast$ and  Eq.~\eqref{eq:kD} is satisfied.

Fig.~\ref{fig:cell} demonstrates the parameter degeneracy in CMB power spectra, which are
computed using {\tt CAMB} \cite{Lewis:1999bs} with recombination code {\tt HyRec} \cite{AliHaimoud:2010dx},
with effects of varying $m_e$ being incorporated in full.
We here vary $m_e$ by $\pm5$\% with $\omega_b$, $\omega_m$ and $h$ being varied simultaneously according to Eqs. \eqref{eq:sim1} and \eqref{eq:sim2}. Except for low-$\ell$ in $C_\ell^{TT}$, where the late-ISW effect is significant, 
CMB power spectra remain remarkably unchanged.
This manifests that varying $m_e$ satisfies all the first three conditions we raised in Introduction.

\begin{figure*}[!t]
\centering
\begin{tabular}{cc}
\includegraphics[width=9cm]{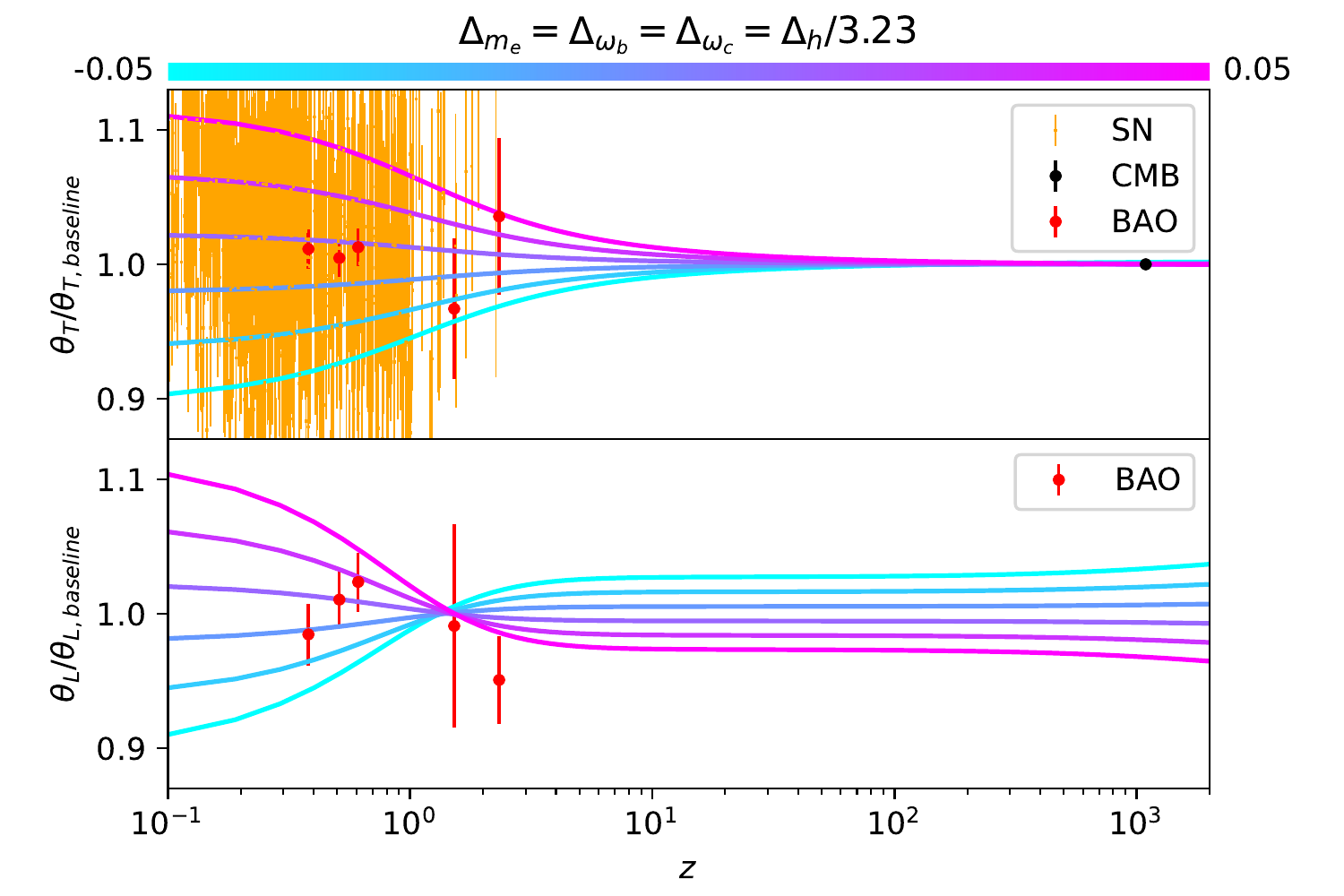} 
&
\includegraphics[width=9cm]{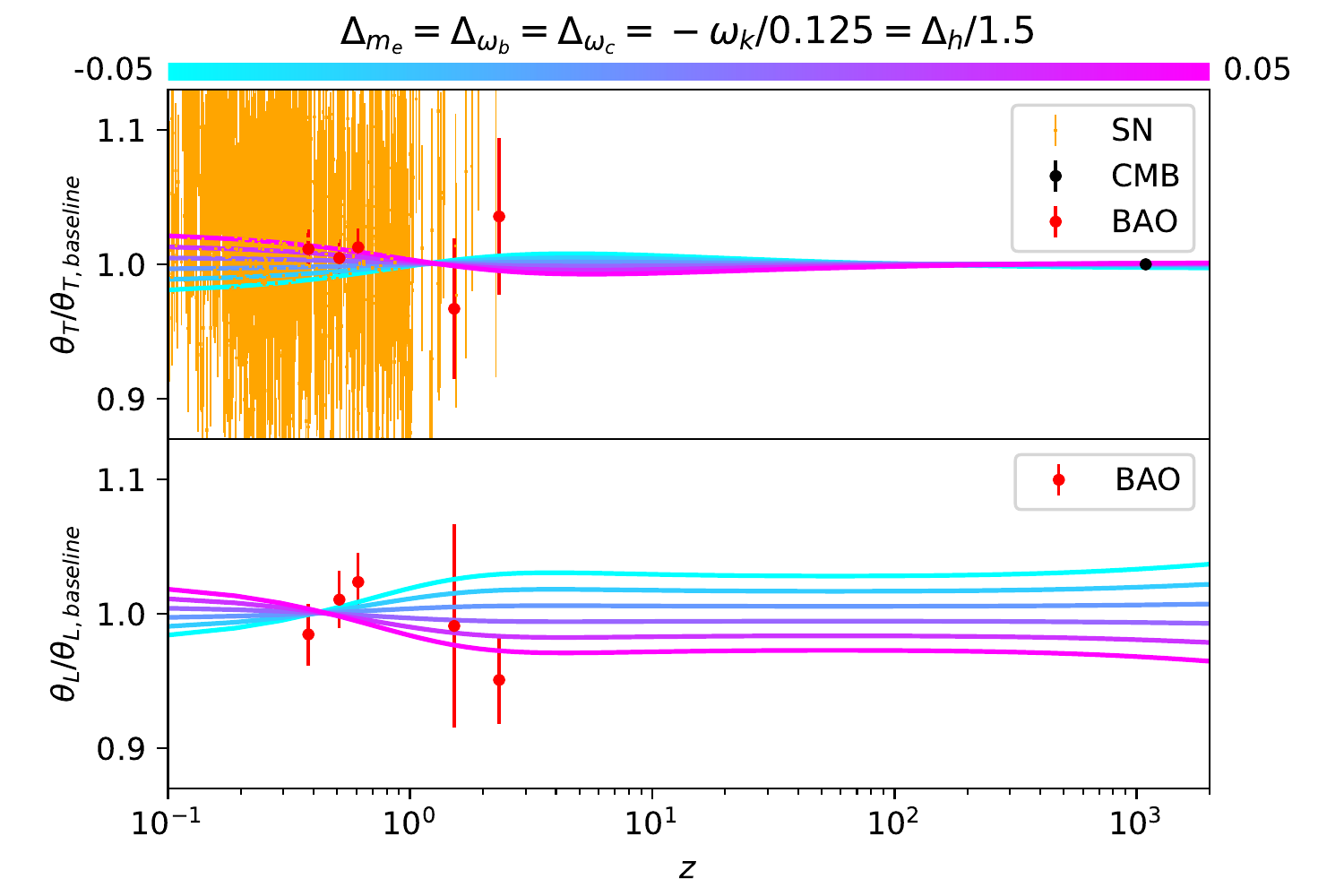}
\end{tabular}
\caption{\label{fig:thetaTL}
(Left) Transverse $\theta_T(z)$ (upper) and longitudinal $\theta_L(z)$ (lower) BAO separations
in varying $m_e$ with $\Lambda$CDM background. Color bar indicates the value of 
$\Delta_{m_e}\in\{-0.05,-0.03,-0.01,0.01,0.03,0.05\}$
each line has. Other cosmological parameters, $(\omega_b,~\omega_c,~h)$, 
are varied with $m_e$ in accordance with Eqs. \eqref{eq:sim1}, \eqref{eq:sim2} and \eqref{eq:me}.
CMB, BAO and (normalized) SNeIa data are also plotted.
(Right) Same as in left panel but 
with $\Omega_k\Lambda$CDM background, with 
Eq. \eqref{eq:sim2} being replaced by Eq. \eqref{eq:sim3}.
}
\end{figure*}

{\it Low-$z$ distances.---}
While CMB spectra are almost conserved, the parameter modification Eqs.~\eqref{eq:sim1} and \eqref{eq:sim2}
in general also modifies late-time expansion and geometric distances, which is severely constrained by BAO and SNeIa.
To see this, we plot the late-time distance and the expansion history in Fig.~\ref{fig:thetaTL}.
Here we have introduced two quantities:
\be
\theta_T(z) \equiv \frac{r_s(z_{\rm drag})}{D_M(z)}, \quad
\theta_L(z) \equiv r_s(z_{\rm drag})[aH](z)
\ee
which are nothing but the scales of BAO measured along the transverse and line-of-sight directions, respectively.\footnote{
Precisely speaking, 
$\theta_L(z)$ is the separation of the BAO scale along the line-of-sight in $\log(1+z)$.
}
In addition, recent BAO \cite{Alam:2016hwk,Zarrouk:2018vwy,Bautista:2017zgn} 
and SNeIa data \cite{Scolnic:2017caz} are overlaid in the same figure.\footnote{
We have normalized the 
SNeIa luminosity distances to give $D_M(z)$ consistent with BAO at $z\simeq0.5$.
}

When $\Lambda$CDM background is assumed (left panel of Fig. \ref{fig:thetaTL}),
the model becomes effectively one-parameter model according to Eqs.~\eqref{eq:sim1} and \eqref{eq:sim2}.
One can see that late-time geometry changes as $m_e$ varies from the baseline.
Since the baseline model fits with BAO data well, $m_e$ is tightly constrained when CMB is combined with BAO, 
and somewhat less tightly but also with SNeIa. 

In the $\Lambda$CDM background, there're no more degrees of freedom to tune the late-time geometry while $\Delta_{m_e}$ is kept nonzero,
and hence it is impossible to solve the $H_0$ tension just with varying $m_e$.
However, it easily becomes possible when the background model is extended appropriately.
In the right panel of Fig.~\ref{fig:thetaTL}, the background cosmology is extended to allow a non-flat Universe ($\Omega_k\Lambda$CDM hereafter)
and plotted is the late-time geometry along a parameter direction
\be
\Delta_h=1.5\Delta_{m_e},\qquad \omega_k = -0.125\Delta_{m_e}, \label{eq:sim3}
\ee
instead of Eq.~\eqref{eq:sim2}. This realizes a good fit to the low-$z$ distance observations even with $\Delta_{m_e}$ as large as 5\%. 
Curvature of the Universe is playing an essential role here. As can be read from Eq.~\eqref{eq:DM}, deviations from flatness grow as 
$\chi(z)\sqrt{|\omega_k|}/L$
 increases. Therefore, 
the curvature selectively affects only the angular diameter distance to CMB and offers the freedom
for low-$z$ and CMB distances to be fitted well simultaneously even with large $\Delta_{m_e}$.
Therefore all the four conditions in Introduction are satisfied in $\Omega_k\Lambda$CDM background with varying $m_e$.

{\it MCMC parameter estimation. ---}
We perform Markov chain Monte Carlo (MCMC) analysis using {\tt CosmoMC} \cite{Lewis:2002ah} 
modified to incorporate varying $m_e$.  We adopt the Planck 2018 reference CMB likelihood TT,TE,EE+lowE \cite{Aghanim:2019ame} in combination with the BAO 
\cite{Alam:2016hwk,Zarrouk:2018vwy,Bautista:2017zgn} and SNeIa data \cite{Scolnic:2017caz}.
We checked that our analysis is consistent with a previous work \cite{Hart:2019dxi} when a $\Lambda$CDM background is assumed.

\begin{figure}[!t]
\centering
\includegraphics[width=7cm]{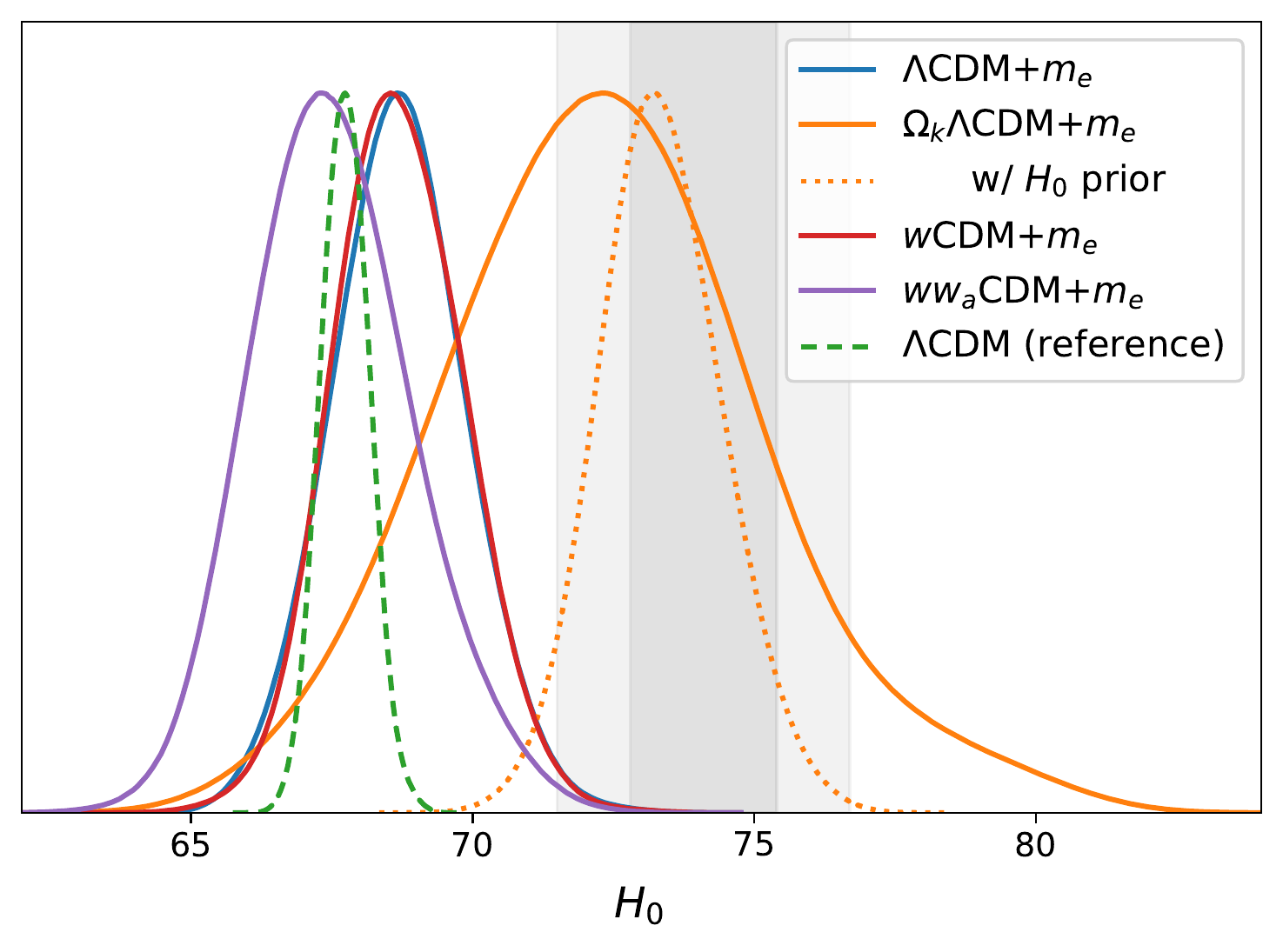} 
\caption{
\label{fig:H0}
Posterior distributions of $H_0$ for varying $m_e$ with different background models and the reference model. 
Gray band shows the direct $H_0$ measurement 
$H_0=74.1\pm1.3\,{\rm km/sec/Mpc}$ without SNeIa \cite{Riess:2020sih}.
Solid and dashed lines are obtained from the combination CMB+BAO+SNeIa.
In order for demonstration, we also depicted the posterior distribution from CMB+BAO+SNeIa+H0
only for varying $m_e$ model with $\Omega_k\Lambda$CDM background (orange dotted line).
}
\end{figure}

Fig.~\ref{fig:H0} shows the posterior distribution of $H_0$ in models with varying $m_e$ in different background including
$\Lambda$CDM, $\Omega_k\Lambda$CDM, $w$CDM, where dark energy (DE) equation of state (EoS) $w$ is assumed to be constant,
and $ww_a$CDM models, where DE EoS is parametrized as in \cite{Chevallier:2000qy,Linder:2002et}.
For reference, $\Lambda$CDM model without varying $m_e$ (``reference" model hereafter) is also plotted. 
We also compare those posterior distributions with the direct measurements $H_0=74.1\pm1.3\,{\rm km/sec/Mpc}$ (Hereafter H0) 
\cite{Riess:2020sih},\footnote{
To minimize influence of systematic errors associated to SNeIa, we here adopt direct $H_0$ measurements without SNeIa.
} which is not
incorporated in the default parameter estimation. 

From the figure, one can immediately see that the varying $m_e$ in $\Omega_k\Lambda$CDM model gives a posterior distribution
matching well with the direct measurements. As expected from the parameter degeneracies discussed above,  
the $\Omega_k\Lambda$CDM background allows substantially broader distributions compared to the reference model.

Besides, it is remarkable that the distribution peak coincides with the direct $H_0$ measurements. The preference for higher $H_0$ in association with $\Omega_k<0$ is brought about by the Planck data at $\ell>30$, which is known to favor larger lens amplitude, $A_{\rm L}>1$ \cite{Aghanim:2018eyx}. Indeed, we found that the posterior mean values in our analysis, which are consistent with local $H_0$ measurements, yield $C^{TT}_\ell$
at $\ell\gtrsim800$ similar to that from the baseline but with $A_{\rm L}=1.1$. Although a closed Universe enhances the CMB lensing effect, in general the fit to BAO and SNeIa gets worse \cite{DiValentino:2019qzk}. However, varying $m_e$ in $\Omega_k\Lambda$CDM model can keep the fit to BAO and SNeIa excellent.

While $H_0$ tension is relaxed with varying $m_e$ in other backgrounds too, as posterior distributions become broader compared to the reference model, their peaks are still displaced from the direct measurements.  This is because these models lack the freedom to fit with the CMB and low-$z$ distances simultaneously.

Table \ref{tab:ppd} summarizes mean values and 68\% intervals of $H_0$ from
the default data set, CMB+BAO+SNeI,a and an extended one, CMB+BAO+SNeIa+H0
as well as the effective $\Delta\chi^2$ for CMB+BAO+SNeIa+H0 
against the reference model. As expected, varying $m_e$ in $\Omega_k\Lambda$CDM model 
yields a significant improvement in data fits with $\Delta \chi_{\rm eff}^2<-23$. 
This is caused not only by resolving the $H_0$ tension, but also by improving the fit to CMB data, 
which alone reduces $\Delta\chi^2_{\rm eff}$ by 6.2.
Our results clearly prove that varying $m_e$ in $\Omega_k\Lambda$CDM is preferred by data over the 
reference model. 

\begin{table*}[!t]
\centering
\caption{\label{tab:ppd} Summary of estimation of $H_0$ and $\Delta\chi^2_{\rm eff}$.
}
\begin{tabular} {lccccc}
\hline\hline
& \multicolumn{4}{c}{$\overbrace{\hspace{5cm}}^{\mbox{varying $m_e$}}$} & constant $m_e$ \\
& $\Lambda$CDM & $\Omega_k\Lambda$CDM & $w$CDM & $ww_a$CDM 
& $\Lambda$CDM (reference) \\
\hline
$H_0$ [km/sec/Mpc]  (mean with 68\% errors) & & & & & \\
\qquad based on CMB+BAO+SNeIa & 
$~68.7_{-1.2}^{+1.2}~$ & $~72.3_{-2.8} ^{+2.7}~$ & $~68.7_{-1.2}^{+1.1}~$ & $~67.5_{-1.6}^{+1.3}~$ & $~67.7_{-0.4}^{+0.4}~$\\
\qquad based on CMB+BAO+SNeIa+H0 & 
$~71.2_{-0.9}^{+0.9}~$ & $~72.9_{-1.0}^{+1.0}~$ & $~71.0_{-1.0}^{+1.0}~$ & $~71.5_{-0.9}^{+1.1}~$ & $~68.4_{-0.4}^{+0.4}~$\\
$\Delta\chi^2_{\rm eff}$ relative to the reference & & & & & \\
\qquad based on CMB+BAO+SNeIa+H0 & $-12.2$ & $-23.5$ & $-12.5$ & $-13.2$ & $0$ \\
\hline\hline
\end{tabular}
\end{table*}

{\it Discussion.---}
The parameter degeneracy in Eq.~\eqref{eq:sim1} is not perfect and varying $m_e$ distorts CMB power spectra through non-equilibrium nature of recombination. Therefore, CMB-S4 \cite{Abazajian:2016yjj} may be able to constrain/verify our examples. Substantial deviations from the baseline in low-$z$ distances are also predicted. For instance, in varying $m_e$ with $\Omega_k\Lambda$CDM background, $\Delta_{r_s(z_{\rm drag})}\simeq -0.05$ and $\Omega_k\simeq-0.01$ are required to solve the $H_0$ tension. Future distance measurements will be able to test such deviations from the baseline \cite{Denissenya:2018zcv}.

\begin{acknowledgements}
This work is supported by JSPS KAKENHI Grant Numbers 18H04339 (TS), 18K03640 (TS), 17H01131 (TT, TS), 19K03874 (TT) and
MEXT KAKENHI Grant Number 19H05110 (TT). This research was conducted using the Fujitsu PRIMERGY CX600M1/CX1640M1 
(Oakforest-PACS) in the Information Technology Center, The University of Tokyo.
\end{acknowledgements}

\pagebreak
\bibliography{h0tension}
\end{document}